\documentclass[twocolumn,showpacs,showkeys,preprintnumbers,amsmath,amssymb]{revtex4}

\usepackage{graphicx}
\usepackage{dcolumn}
\usepackage{bm}
\usepackage{amsfonts}

\begin{document}

\title{Account of Nuclear Scattering at Volume Reflection}

\author{M.V. Bondarenco}
\email{bon@kipt.kharkov.ua}
\affiliation{%
Kharkov Institute of Physics and Technology, 1 Academic St., 61108
Kharkov, Ukraine
}%

\date{\today}

\begin{abstract}
For a particle traversing a bent crystal in the regime of volume
reflection we evaluate the probability of interaction with atomic
nuclei. Regardless of the continuous potential shape, this
probability is found to differ from the corresponding value in an
amorphous target by an amount proportional to the crystal bending
radius, and the particle deflection angle. Based on this result, we
evaluate the rate of inelastic nuclear interactions, and the final
beam angular dispersion due to multiple Coulomb scattering. The
theoretical predictions are compared with the experiments. The
impact of multiple Coulomb scattering on the mean volume reflection
angle is also discussed.
\end{abstract}

\pacs{61.85.+p, 29.27.-a, 45.10.-b}
\keywords{bent crystal; volume reflection; inelastic nuclear
interactions; multiple Coulomb scattering}

\maketitle



\section{Introduction}

Deflection of fast charged particle beams by bent crystals is a
developing branch of accelerator technology, pursuing the goal of
steering ultra-high energy beams in a restricted laboratory space.
Till the last decade, the conventionally exploited deflection
mechanism in bent crystal tools was channeling
\cite{ref:chann-defl}, but an alternative promising technique based
on volume reflection \cite{ref:VR-technology} has emerged recently,
dealing with over-barrier particles, and offering the benefit of
large angular acceptance (equal to the crystal total bending angle)
and almost 100\% deflection efficiency. Some drawback of the volume
reflection technique is the smallness of the deflection angle, which
is of the order of Lindhard's critical angle
$\theta_c=\sqrt{2V_0/E}$, with $V_0$ the planar potential well
depth, and $E\ggg V_0$ the particle energy. This angle is
independent of the crystal thickness, but a buildup of the
deflection angle can nonetheless be acheived by transmitting the
particle through a sequence of bent crystals \cite{MVR}, or
arranging a composite volume reflection from several atomic planes
in one crystal \cite{Composite-VR}. An additional advantage of the
volume reflection mechanism is that it applies equally well to
negative particles, for which channeling is spoiled by the
unavoidable multiple Coulomb scattering on atomic nuclei residing at
the bottom of the potential well.

The origin of the volume reflection effect is essentially dynamical,
due to non-linearity of the motion in the strong continuous
potential of bent atomic planes. At that, incoherent scattering
effects on individual atomic nuclei can be held small compared to
the deflection angle mean value. That even admits using for volume
reflection experiments the crystals of thickness $1\div2$ mm, which
by $\sim10$ times exceeds the volume reflection region extent,
estimated as $\sim R\theta_c$ \cite{ref:Tar-Scand}.\footnote{At
typical beam energy $E\simeq400$ GeV (CERN SPS) entailing
$\theta_c\sim10^{-5}$ rad, and at optimal radius $R\sim10$ m, this
longitudinal scale amounts to $R\theta_c\sim10^{-1}$ mm. } However,
in other respects, for instance for evaluation of the deflected beam
angular spread, the multiple Coulomb scattering can not be ignored.
Worth mentioning also are the inelastic nuclear interactions, which
can be registered as multiple secondary particle signals in the beam
loss monitors. Such events are relatively rare, and therefore
present no danger for the primary beam propagation, but from their
rate one can determine the robustness of the volume reflection
inside the crystal. In fact, since physically they occur due to
interaction with the same nuclei as Coulomb scattering, the rates of
these processes must be closely related. Similar processes also
include atomic K-shell ionization and the accompanying
characteristic X-ray radiation occurring at fast charged particle
passage close to an atomic nucleus.

Outside the volume reflection area, we know that the particle motion
becomes highly over-barrier, whereby the rate of fast particle
scattering on atomic nuclei must approach that in an amorphous
medium. Thus, in the thick-crystal limit, the number of nuclear
interactions in the whole crystal roughly equals that in an
amorphous target of same material and thickness. But that kind of
approximation is unsuitable if the crystal thickness is comparable
with the volume reflection area extent, or if one aims to exploit
the nuclear interactions, for monitoring the in-depth particle
dynamics. Since the number of nuclear interactions in an amorphous
medium, is known precisely enough from various experiments, it may
be neatly subtracted and in this way the difference be measured. Yet
the theory interpreting this difference needs to be developed.

For a consistent treatment of incoherent multiple scattering at
volume reflection, one needs to solve the kinetic equation in a
non-uniform external field, but in general that does not seem
feasible by analytic means. Still, there must exist a domain in
which the Coulomb scattering is sufficiently small for perturbative
account, and there are reasons to expect it to cover the practically
interesting range of beam and crystal parameters where most of the
presently available data belong.

The aim of the present paper is to evaluate the rate of nuclear
interactions adhering to the framework of the perturbative approach.
Given the 1d (radial) character of volume reflection dynamics in the
pure continuous potential of a uniformly bent crystal, the
corresponding particle trajectories may be expressed analytically,
and the probability of a nuclear interaction be calculated along the
known path. In Sec.~\ref{sec:DeltaL} we implement this procedure,
for particles of different charge sign, and different orientations
of the crystal: (110) and (111). In Sec.~\ref{sec:Rmult} limitations
of the perturbative approach are determined. In Sec.~\ref{sec:Inel}
we apply the results of Sec.~\ref{sec:DeltaL} to the probability of
inelastic nuclear interactions at specific experimental conditions,
comparing the theory with the experiment. In Sec.~\ref{sec:spread}
we extend the theory predictions to elastic scattering and address
the issue of the volume-reflected beam angular divergence. The
conclusions are given in Sec.~\ref{sec:Summary}.




\section{Inelastic nuclear interaction probability under neglect of Coulomb
scattering}\label{sec:DeltaL}

Volume reflection process assumes particle interaction with the bent
crystal in a planar orientation. At that, the atomic density in each
plane may be regarded as uniform. Since all the nuclei are located
in the planes, one particle crossing of an atomic plane may be
viewed as an elementary act of nuclear interaction.
If all the planes are packed them with the same density and
equidistant\footnote{For a silicon crystal, that corresponds to
planar orientaton (110), which is most frequently used in
experiments on volume reflection. Another used case is planar
orientation (111), involving 2 non-equidistant planes within the
period. Extension to that case is straightforward and will be
considered later in this section.}, with the inter-planar distance
$d$, the probability of any kind of nuclear interaction in one
atomic plane crossed at a tangential angle $\theta$ is
\begin{equation}\label{P1}
    P_1=n_{\mathrm{at}}\sigma_A \frac{d}{\sin\theta},
\end{equation}
where $\sigma_A$ is the corresponding cross-section on a single
nucleus, and $n_{\mathrm{at}}$ the atomic density in the crystal
volume. For elastic scattering one must employ the transport
cross-section ($\sigma_A=\sigma_{\mathrm{tr}}$), while for inelastic
interactions -- the total inelastic cross-section on a Silicon
nucleus ($\sigma_A=\sigma_{\mathrm{inel}}$). More specifically, the
inelastic interactions may be genuinely nuclear, involving multiple
hadron production, or electromagnetic, e.g., ionization of the
small-radius atomic K-shell\footnote{For silicon ($Z=14$), the
atomic K-shell radius is $r_B/Z\sim 0.03{\AA}$, which is quite small
compared to the interplanar distance $d\sim2\AA$. To knock out a
K-shell electron, a relativistic charged particle must pass at a
distance from the electron yet much smaller than $r_B/Z$. Thus, to
ionize the K-shell, the charged particle has to pass through a
really close vicinity of the atomic nucleus, and in this respect the
process is similar to a nuclear interaction.} and the accompanying
characteristic X-ray radiation. In the Glauber approximation, all
the high-energy nuclear and electromagnetic cross-sections are
energy-independent, though one might take their weak energy
dependence into account, if desired.

In the case of a straight crystal, and for a highly over-barrier
particle, when $\theta$ is much greater than the critical value,
\begin{equation}\label{high-overbarrier}
    \theta\gg\theta_c,
\end{equation}
and thus is subject to negligible variation within the crystal
$(\mathrm{var}\theta\lesssim\theta_c\ll\theta)$, summing up
contributions (\ref{P1}) for $\approx\frac{L\sin\theta}{d}$ crossed
planes would yield the total nuclear interaction probability:
\begin{equation}\label{P-amorph}
    P=n_{\mathrm{at}}\sigma_A L\qquad \left(
                                                                                    \begin{array}{c}
                                                                                      \text{straight
    crystal}, \\
                                                                                      \text{highly over-barrier motion} \\
                                                                                    \end{array}
                                                                                  \right).
\end{equation}
This value is independent of $d$ and $\theta$, and is equal to the
corresponding probability in an amorphous medium -- not surprisingly
since the particle flow covers each nucleus with the same density,
equal to that in the initial beam outside the crystal. In this
sense, one can speak about an ``amorphous orientation" of a perfect
crystal even. By the same token, the latter notion applies in bent
crystal regions where the particle motion is highly over-barrier.

In the volume reflection case, however, condition
(\ref{high-overbarrier}) in the vicinity of a radial reflection
point breaks down. In this region the plane crossing angle varies
considerably along the particle path, hence it is to be evaluated
accurately at each plane crossing. To this end, the particle
trajectory must be computed beyond the straight line approximation
and, neglecting the multiple scattering, we are going to approximate
it by a trajectory in the pure continuous potential. If the crystal
is bent uniformly (which is sufficiently credible nowadays), the
trajectory description simplifies in cylindrical coordinates, with
the bent planes corresponding to surfaces of constant radius
relative to some axis far outside of the crystal. Thereat, the plane
crossing angle sine entering Eq. (\ref{P1}) expresses simply as the
time derivative of the particle radial coordinate:
\begin{equation}\label{sin}
    \sin\theta\approx \dot r.
\end{equation}
Inserting Eq.~(\ref{sin}) to Eq.~(\ref{P1}) and summing over all the
planes crossed by the particle, we obtain the total inelastic
nuclear interaction probability in a bent crystal:
\begin{equation}\label{P}
    P\approx n_{\mathrm{at}} \sigma_A d\sum_n\frac{1}{\dot r_n}\qquad (\mathrm{in\, a\, uniformly\, bent\,
    crystal}).
\end{equation}

Since the particle motion straightens out away from the volume
reflection area, there the nuclear interaction rate per unit length
should approach that in an amorphous medium. Hence, the difference
between the number of nuclear interactions in an oriented crystal
and in an ``unoriented" crystal may be expressed as
\begin{equation}\label{via}
    \Delta P=n_{\mathrm{at}}\sigma_A\Delta L,
\end{equation}
where the isolated geometrical factor
\begin{equation}\label{L-difference}
    \Delta L=\lim_{L\to\infty}\left(\sum_n\frac{d}{\dot
    r_n}-L\right)
\end{equation}
is expected to be finite and independent of $L$, representing the
excess (or deficit) of the target nuclear interaction range.

\begin{figure}
\includegraphics{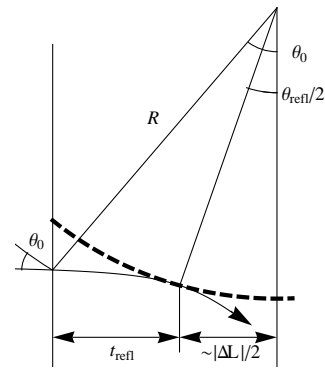}
\caption{\label{fig:angle-diff} Relation between the volume
reflection angle, the angle of particle entrance to the crystal, the
depth of the reflection point and the crystal bending radius. Dashed
arcs -- bent atomic planes. Solid curve -- schematic of the particle
trajectory.}
\end{figure}

In this section we will concentrate on evaluation of limit
(\ref{L-difference}). To begin with, note that the trajectory in a
centrally-symmetric continuous potential is symmetric with respect
to the reflection point, enabling one to count the crossed planes
beginning from the reflection point in only one direction and double
the result:
\begin{equation}\label{difference}
    \Delta L=2\lim_{n_{\max}\to\infty}\left(\sum_{m=0}^{n_{\max}}\frac{d}{\dot
    r_{n_{\max}-m}}-t_{\mathrm{refl}}\right).
\end{equation}
Here $t_{\mathrm{refl}}$ is the distance from the reflection point
to the crystal boundary -- say, its front face, where the entrance
angle $\theta_0$ relative to the atomic planes is known (see
Fig.~\ref{fig:angle-diff}). This quantity is expressible through the
particle trajectory, too. In fact, at large $n_{\max}$ it is
unambiguously related with the total volume reflection angle
$\theta_{\mathrm{refl}}$, which can be evaluated in a model approach
\cite{ref:Bond-VR}, or measured. In the small-angle approximation,
from Fig.~\ref{fig:angle-diff} one infers\footnote{In this article,
we choose the angle counting direction so that volume reflection
angles (opposite in sign to the crystal bending direction) were
positive. In the literature it is more conventional to choose the
channeling angles positive, while volume reflection angles are then
negative. But in our problem we only deal with the volume reflection
angles, which moreover enter to the observable $\Delta L$ linearly,
so we permit ourselves to alter the convention.},
\begin{equation}\label{Rtheta}
    \theta_{\mathrm{defl}}/2=\lim_{\theta_0\to\infty}(\theta_0-t_{\mathrm{refl}}/R).
\end{equation}

Having traded $t_{\mathrm{refl}}$ for $\theta_0$, the latter angle
now is to be related with $n_{\max}$. But this relation, in fact,
appears to be trivial in the thick crystal limit whose condition is
\begin{equation}\label{}
    \theta_0^2=\frac{L^2}{R^2}\gg\frac{V_0}{E}\qquad (\text{``thick"-crystal limit}).
\end{equation}
Indeed, utilizing the transverse energy conservation law in a
cylindrically symmetric continuous potential $V(r)$,
\begin{equation}\label{Eperp}
    E_{\perp}=\frac E2\dot{r}^2+V(r)-E\frac{r}{R},
\end{equation}
we extract
\begin{equation}\label{nm}
    n_{\max}=\frac{R\theta^2_0}{2d}+\mathcal{O}\left(\frac{VR}{Ed}\right)\equiv\frac{R\theta^2_0}{2d}+\mathcal{O}\left(\frac{R}{R_c}\right),
\end{equation}
where
\begin{equation}\label{Rc}
    R_c\simeq\frac{Ed}{4V_0}\approx \mathrm{cm}\frac{E}{5 \, \mathrm{GeV}}
\end{equation}
is the Tsyganov's critical radius \cite{ref:chann-defl}. Solving
Eq.~(\ref{Rc}) for $\theta_0$, we get
\begin{equation}\label{}
    \theta_0=\sqrt{\frac{2d}{R}}\sqrt{n_{\max}}+\mathcal{O}\left(\sqrt{\frac{R}{2dn_{\max}}}\frac{V}{2E}\right).
\end{equation}
Here the interaction-dependent correction term asymptotically
vanishes as $n_{\max}\to\infty$, and may be omitted under the limit
sign.

Combining Eqs.~(\ref{difference}, \ref{Rtheta}) and (\ref{nm}), we
cast $\Delta L$ in form
\begin{equation}\label{DL}
    \Delta L=R\theta_{\mathrm{defl}}+2\sqrt{2Rd}\lim_{n_{\max}\to\infty}\left(\sum_{m=0}^{n_{\max}}\frac{\sqrt{d/2R}}{\dot
    r_{n_{\max}-m}}-\sqrt{n_{\max}}\right).
\end{equation}

\subsection{Orientation (110)}

To proceed, we need to evaluate the terms of the sequence of angles
$\dot{r}_n$ entering the denominator in Eq.~(\ref{DL}). For a
silicon crystal in planar orientation (110), $\dot{r}_n$ actually
appears to be a fairly simple function of the plane order number
$n$, independently of the precise shape of the inter-planar
continuous potential. From Eq.~(\ref{Eperp}) we express
\begin{equation}\label{rdot}
    \dot r=\sqrt{2\left(\frac{E_\perp-V}E+\frac{r}R\right)},
\end{equation}
with $E_\perp=E_\perp(\theta_0,b)$ depending on the particle initial
conditions including its impact parameter $b$ and the incidence
angle $\theta_0$ with respect to the planes. Now, granted the
periodocity of the intra-crystal continuous potential, values of
$V(r)$ are equal at atomic plane locations:
\begin{equation}\label{}
    V(r_n)=V|_{r\in \mathrm{at. plane}}=\mathrm{const}.
\end{equation}
As for $r$ in the centrifugal energy term in Eq.~(\ref{rdot}), its
value at different atomic planes differs only by a multiple of $d$:
\begin{equation}\label{}
    r_n=r_{n_{\max}}+(n_{\max}-n)d \qquad (n\leq n_{\max}).
\end{equation}
With this in mind, we can write
\begin{equation}\label{r-m}
    \dot r_{n_{\max}-m}=\sqrt{2\frac dR(\eta+m)},\quad m=0,1,2,... ,
\end{equation}
where variable
\begin{equation}\label{xi-def}
    \eta=\frac{(E_{\perp}-V(r_{n_{\max}}))R}{Ed}+\frac{r_{n_{\max}}}{d}
\end{equation}
accumulates all the dependence on the inital conditions. An
important notice is that it belongs to an interval of unit length:
\begin{equation}\label{}
    \eta_{\min}<\eta\leq\eta_{\min}+1,
\end{equation}
in order to secure the relation $\min\{\dot r^2_n\}=\max\{\dot
r^2_{n-1}\}$.

\begin{figure}
\includegraphics{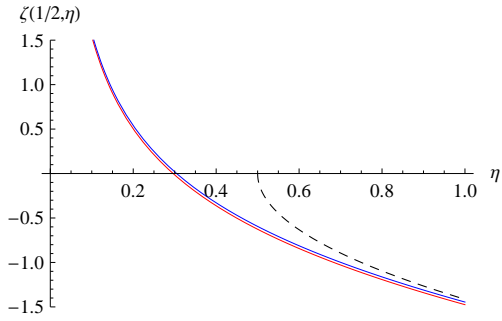}
\caption{\label{fig:zeta} Blue curve -- behavior of Hurwitz zeta
function (\ref{zeta-defin}), red curve -- its approximation
(\ref{zeta-approx-forall}), black dotted curve -- approximation
(\ref{zeta-approx-foreta>1}) valid for $\eta>1$. The integral from
the function over this interval equals zero (Eq.~(\ref{int=0})).}
\end{figure}

Substituting $r_{n_{\max}-m}$ from Eq.~(\ref{r-m}) to
Eq.~(\ref{DL}), we cast it in form
\begin{equation}\label{DelL}
    \Delta
    L=R\theta_{\mathrm{defl}}+\sqrt{2Rd}\zeta\left(\frac12,\eta\right)\qquad (\mathrm{Si}\,(110)),
\end{equation}
where
\begin{equation}\label{zeta-defin}
    \zeta\left(\frac12,\eta\right)=\lim_{n_{\max}\to\infty}\left(\sum_{m=0}^{n_{\max}}\frac{1}{\sqrt{\eta+m}}-2\sqrt{n_{\max}}\right).
\end{equation}
The latter function is categorized as Hurwitz (or generalized
Riemann) zeta with the parameter equal $\frac12$ (for a general
definition of $\zeta(\alpha,v)$ see \cite{ref:zeta}). For all
practical purposes, it may be approximated by
\begin{equation}\label{zeta-approx-forall}
    \zeta\left(\frac12,\eta\right)\approx\frac1{\sqrt{\eta}}+\frac1{2\sqrt{\eta+1}}-2\sqrt{\eta+1},\quad
    \forall \eta,
\end{equation}
obtained by application to the sum in (\ref{zeta-defin}) (with the
first, singular term being singled out) of the Euler-Maclaurin
formula \cite{Eul-Macl}. Furthermore, at $\eta>1$ it admits a
simpler approximation
\begin{equation}\label{zeta-approx-foreta>1}
    \zeta\left(\frac12,\eta\right)\approx-2\sqrt{\eta-\frac12}\qquad (\eta>1)
\end{equation}
(see Fig.~\ref{fig:zeta}).

\subsection{Orientation (111)}

In case of orientation (111), the continuous potential values at all
the planes are still equal, but the intreplanar intervals assume
alternating values $d/4$ and  $3d/4$. Thereat, the calculation
principle remains the same, except that the summation over the atomic
planes is carried out separately for odd and even numbers. The
result then involves two different $\zeta$ functions:
\begin{equation}
\Delta L=
R\theta_{\mathrm{defl}}+\sqrt{\frac{Rd}{2}}\left[\zeta\!\left(\frac12,\eta_1\right)+\zeta\!\left(\frac12,\eta_2\right)\!
\right]\quad (\mathrm{Si}\, (111)),
\end{equation}
where depending on which of the non-equivalent planes is encountered
the last,
\begin{subequations}
\begin{equation}\label{prob34}
\eta_{\min}<\eta_1\leq\eta_{\min}+\frac34, \quad
\eta_2=\eta_1+\frac14,
\end{equation}
or
\begin{equation}\label{prob14}
\eta_{\min}<\eta_1\leq\eta_{\min}+\frac14, \quad
\eta_2=\eta_1+\frac34.
\end{equation}
\end{subequations}
For positive particles, the probability of case (\ref{prob34})
equals $3/4$, while that of case (\ref{prob14}) is $1/4$. For
negative particles at $R>4R_c$ the case (\ref{prob34}) is realized
with the unit probability, because the higher potential barriers
completely shadow the minor ones, even in spite of the centrifugal
energy tilt.

So far the formulation held for particles of any charge sign. But in
what concerns the last unknown quantity $\eta_{\min}$, the situation
turns principally different for positively and for negatively
charged particles. Below we will analyze these two cases separately.

\subsection{Positively charged particles}

\begin{figure}
\includegraphics{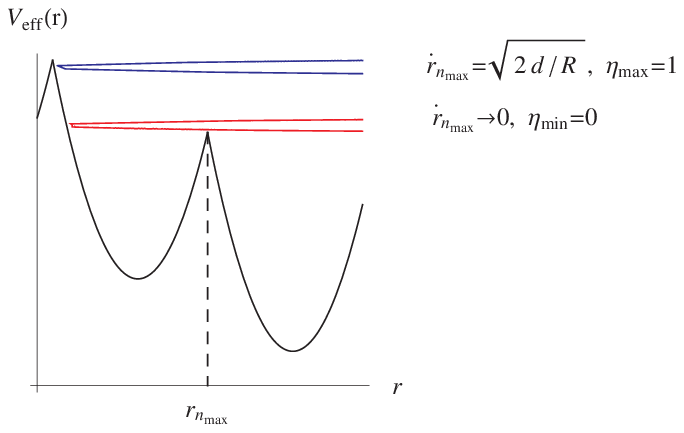}
\includegraphics{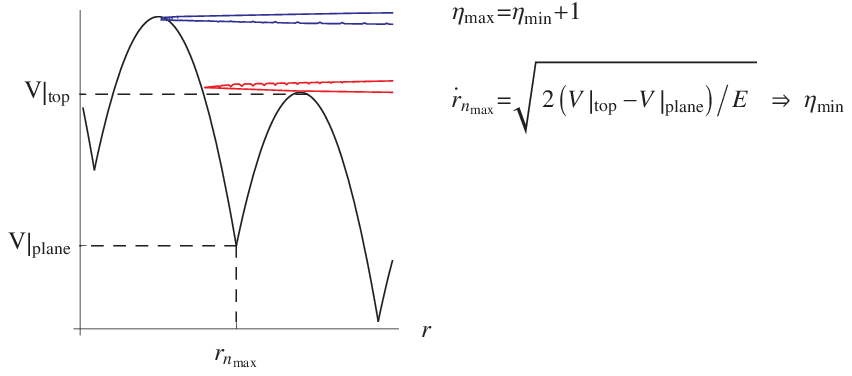}
\caption{\label{fig:potential} Determination of the range of
variation of parameter $\eta$ in crystal orientation (110); (a) --
for positively charged particles; (b) -- for negatively charged
particles. For details see text.}
\end{figure}

In the case of positively charged particles (typically protons,
which are most important for ultra-high energy accelerator
applications), determination of $\eta_{\min}$ is particularly
simple. Consider, again, the case of orientation (110). At particle
entrance to the reflection interval (see Fig.~\ref{fig:potential}a),
the minimum of the kinetic energy is achieved when the particle
passes the last potential barrier with a vanishing kinetic energy. But
for positively charged particles that barrier coincides with the
atomic plane, the kinetic energy on which we want to know. So, at
$\eta_{\min}$ this energy merely tends to zero:
\begin{equation}\label{etamin-pos}
    \min\left\{\dot r^2_{n_{\max}}\right\}=0 \quad \Rightarrow\quad
    \eta_{\min}=\frac{R}{2d}\min\left\{\dot r^2_{n_{\max}} \right\}=0,
\end{equation}
i.e., $\eta$ belongs to the interval $0<\eta\leq 1$ (actually
exhibited in Fig.~\ref{fig:zeta}). The divergence of function
$\zeta(1/2,\eta)$ at the physical interval end-point $\eta\to0$
corresponds to a grazing crossing of the last atomic plane. The
other end-point value equals $\zeta(1/2,1)=\zeta(1/2)\approx-1.46$.
Note that with the known lower limit (\ref{etamin-pos}), $\eta$ may
even be found without the need to evaluate $n_{\max}$, if one rewrites
Eq.~(\ref{xi-def}) as
\begin{equation}\label{eta-fract}
    \eta\equiv\left\{\eta\right\}_{\mathrm{f}}=\left\{\frac{(E_{\perp}-V|_{r\in \mathrm{at. plane}
    })R}{Ed}+\frac{r_{\mathrm{plane}}}{d}\right\}_{\mathrm{f}},
\end{equation}
with braces $\{\,\}_{\mathrm{f}}$ signifying the operation of taking
the fractional part, and $r_{\mathrm{plane}}$ being $r$ at any
atomic plane (without a difference under the $\{\,\}_{\mathrm{f}}$
sign).

Since at $\eta\lesssim1$ typical values of function $\zeta$ are
$\sim1$, the ratio of the two terms in (\ref{DelL}) is of the order
\begin{equation}\label{rat}
    \frac{\sqrt{2Rd}}{R\theta_c}=\sqrt{\frac{R_cF_{\max}d}{RV_0}}\simeq\sqrt{\frac{4R_c}{R}}.
\end{equation}
It is known that the quality volume reflection is possible provided
$R>4R_c$ (see \cite{ref:Bond-VR}), whereat ratio (\ref{rat}) appears
to be $<1$. But in actual practice that ratio may still be sizeable,
especially at $\eta\to0$ where $\zeta$ function blows up. Then, it
might seem that $\eta$ must necessarily be specified issuing from
the initial conditions.

However, one should remember that in a real beam  the initial
conditions for the particle entering the crystal are not quite
certain. The beam transverse dimensions are always much greater than
inter-atomic distances in the crystal, therefore the particle impact
parameters $b$ have to be averaged over. Besides that, the
indeterminancy $\delta\theta_0$ of the angles in the incident beam
is usually large compared to $d/L$, and needs averaging in a
vicinity of the mean value $\theta_0$.

Examining Eq.~(\ref{eta-fract}), we see that the expression under
the fractional part sign contains contributions from the initial
kinetic and potential energies in an additive way: $E_{\perp}=\frac
E2 \theta^2_0+V(b)$. Actually, fluctuations of both contributions
are large:
\begin{equation}\label{var-theta0}
    \delta\left(\frac{R}{Ed}\frac E2 \theta^2_0\right)\simeq
    \frac{R\theta_0}{d}\delta\theta_0\sim\frac{L}{2d}\delta\theta_0\gg
    1,
\end{equation}
and
\begin{equation}
    \delta\left(\frac{R}{Ed}V\right)\sim\frac{RV_0}{Ed}\sim\frac{R}{4R_c}>1.
\end{equation}
Thus, at fluctuations of particle parameters in a real beam, $\eta$
spans its unit definition interval many times, and each time its
relation with $\theta_0$ and $b$ is approximately linear, so in
fact, $\eta$ may be treated as a uniformly distributed random
variable. Then, averaging over the beam is equivalent to unweighted
averaging over $\eta$. But such an average of function $\zeta$
entering (\ref{DelL}) gives zero due to identity
\begin{equation}\label{int=0}
    \int_0^1d\eta\zeta\left(\frac12,\eta\right)=0,
\end{equation}
straightforwardly checkable from definition (\ref{zeta-defin}). That
leads to a simple and model-independent relation:
\begin{equation}\label{DL2}
    \left\langle\Delta L\right\rangle=R\left\langle\theta_{\mathrm{defl}}\right\rangle\qquad (\text{positively charged particles}).
\end{equation}

The same result (\ref{DL2}) is obtained for orientation (111).

\subsection{Negatively charged particles}

The distinction of the negatively charged particle (typically,
$\pi^-$) case is that atomic plane positions do not coincide with
the tops of the potential barriers -- thus, here $\eta_{\min}>0$
(see Fig.~\ref{fig:potential}b). To determine the value of
$\eta_{\min}$, we need to know the particle kinetic energy at the last plane
crossing, which in the present case equals the difference of
potential energies between the atomic plane and the top of the
preceding barrier:
\begin{equation}\label{}
    \frac E2 \min \dot r^2_{n_{\max}}=V|_{r\in \mathrm{top}}-V|_{r\in
    \mathrm{plane}}\approx V_0\left(1-\frac{R_c}{R}\right)^2.
\end{equation}
Here the last equality is exact for a parabolic inter-planar
potential, while for a non-parabolic one it also turns to be exact
in the limits $R\gg R_c$ and $R\to R_c$, so heuristically we may
expect it to be sufficiently accurate on the whole interval $R>R_c$.
Therewith, we derive
\begin{equation}\label{etamin-neg}
    \eta_{\min}=\frac{R}{2d}\min\dot r^2_{n_{\max}}\approx\frac{V_0R}{Ed}\left(1-\frac{R_c}{R}\right)^2.
\end{equation}
In this expression, $\frac{V_0R}{Ed}\approx\frac{R}{4R_c}$, which
for quality volume reflection is supposed to be $>1$. From
Fig.~\ref{fig:zeta} we see that at $\eta>1$ function
$\zeta(1/2,\eta)$ is fairly smooth and may be linearized in $\eta$
about the midpoint $\eta_{\min}+1/2$. Taylor-expanding
Eq.~(\ref{zeta-approx-foreta>1}), we get:
\begin{equation}\label{zeta-lin}
    \zeta\left(\frac12,\eta\right)\approx-2\sqrt{\eta_{\min}}-\frac1{\sqrt{\eta_{\min}}}\left(\eta-\eta_{\min}-\frac12\right).
\end{equation}

Averaging of the last term of Eq.~(\ref{zeta-lin}) over the unit
interval of $\eta$ gives zero. Substituting in the first term of
(\ref{zeta-lin}) $\eta_{\min}$ from Eq.~(\ref{etamin-neg}), and all
that to Eq.~(\ref{DelL}), we obtain the final expresson for the
average nuclear range difference in the case of (110) orientation:
\begin{eqnarray}\label{DL-neg}
    \left\langle\Delta
    L\right\rangle=R\left\langle\theta_{\mathrm{defl}}\right\rangle-2R\theta_c\left(1-\frac{R_c}{R}\right)\quad\\
    (\text{negatively charged particles, Si(110)}).\nonumber
\end{eqnarray}
At $R\gg4R_c$, when
$\left\langle\theta_{\mathrm{defl}}\right\rangle\approx\theta_c$,
Eq.~(\ref{DL-neg}) reduces to
\begin{equation}\label{39}
    \left\langle\Delta
    L\right\rangle\approx-R\theta_c.
\end{equation}

For orientation (111) in the negative particle case the calculation
is more complicated. We will qoute the result under the condition
$R>4R_c$, retaining only the linear correction in $R_c/R$, which is
relatively simple:
\begin{eqnarray}\label{DL-neg111}
    \left\langle\Delta
    L\right\rangle=R\left\langle\theta_{\mathrm{defl}}\right\rangle-2R\theta_c\left(1-\frac{5R_c}{3R}\right)\quad\\
    (\text{negatively charged particles, Si(111)}).\nonumber
\end{eqnarray}
Here we let $\theta_c=\sqrt{2V_L/E}$, and
$R_c\simeq\frac{3Ed}{16V_L}$, with $V_L$ the larger well depth in a
straight crystal (see \cite{ref:chann-defl}).

Eqs.~(\ref{DL2}, \ref{DL-neg}) are our main results. Comparing them,
we see that for positive particles the nuclear range excess is
positive, while for negative particles it is negative (i. e. there
is a deficit), being of the same order in magnitude. This situation
is rather natural since positive particles are repelled from the
atomic planes and cross them more tangentially, while negative
particles are attracted, crossing the planes more steeply. From the
practical side, it is worth noting that in order to detect  the
discussed effect, it might actually be easier to measure the
difference not between the volume reflection case and the amorphous
orientation, but between volume reflection cases for positively and
negatively charged particles in the same bent crystal -- since
thereat the signal is $\sim1.5$ times larger.

\section{Violation of volume reflection dynamics by multiple scattering}\label{sec:Rmult}

The perturbative treatment of multiple scattering in the previous
section permitted us to derive simple formulae, but their physical
reliability yet depends on the crystal and the beam parameters. In
this section, we will work out quantitative conditions at which the
perturbative treatment of multiple scattering is justified,
concentrating on positive particle case (which is simpler and more
important for applications). We will also touch upon the general
trend in the $\theta_{\mathrm{defl}}$ dependence on $R$ due to
multiple scattering.

\subsection{Condition of coherent dynamics dominance}
The quantity representing the multiple scattering influence on the
volume reflection is the multiple scattering mean square angle
$\sigma_{\mathrm{am}}$ within the volume reflection area extent
$\sim R\theta_{\mathrm{defl}}$. It competes with the dynamical
angles, but it is a subtle issue to find among them the most
sensitive quantity. For positively charged particles, most
certainly, the smallest angle is that of the last atomic plane
crossing, $\dot{r}_{n_{\max}}$. Employing Eq.~(\ref{r-m}), the
average $\dot{r}_{n_{\max}}$ may be estimated as
\begin{equation}\label{aver-r-dot}
    \left\langle\dot{r}_{n_{\max}}\right\rangle=\int_0^1 d\eta \dot{r}_{n_{\max}}=\frac23\sqrt{\frac{2d}{R}}\approx\sqrt{\frac{d}{R}}.
\end{equation}
Note that in contrast to $\theta_c$, it does not depend on the
particle energy, instead involving the crystal bending radius. At
particle energies and crystal radii suitable for volume reflection,
$\left\langle\dot{r}_{n_{\max}}\right\rangle$ is smaller than
$\theta_c$, as long as
\begin{equation}\label{}
    \frac{\left\langle\dot{r}_{n_{\max}}\right\rangle}{\theta_c}\sim\sqrt{\frac{2R_c}{R}}\ll1.
\end{equation}

The angle of multiple scattering (in projection onto \emph{one}
relevant transverse direction perpendicular to the active family of
atomic planes) is determined by the Highland-Lynch-Dahl formula
\cite{ref:Dahl-Lynch} \footnote{This formula was established based
on experiments with hadronic projectiles, which we are exactly
interested in for volume reflection applications. In fact, for
hadronic projectiles nuclear elastic (diffractive) scattering
contributes to the angular diffusion as well, but still in the
literature the whole process is called multiple Coulomb scattering.}
\begin{equation}\label{Dahl-Lynch}
    \sigma_{\mathrm{am}}(T)=\frac{13.6\,\mathrm{MeV}}{E}\sqrt{\frac{T}{X_0}}\left(1+0.038\ln\frac{T}{X_0}\right),
\end{equation}
with $T$ the traversed material thickness, and $X_0$ the
material-dependent radiation length constant (for silicon
$X_0\approx9.36$~cm). The origin of the logarithm of $T$ in
Eq.~(\ref{Dahl-Lynch}) is due to the Rutherford large-angle ``tail"
of multiple scattering, slightly affecting the Gaussianity of the
profile. Formula (\ref{Dahl-Lynch}) works with an accuracy of a few
percent for high- and intermediate-$Z$ substances. At
$T\sim0.2\div1$ mm, i. e. $1+0.038\ln\frac{T}{X_0}\simeq0.8\pm0.03$,
Eq.~(\ref{Dahl-Lynch}) may be used in the simplified form
\begin{equation}\label{sigma-am}
    \sigma_{\mathrm{am}}(T)\approx\frac{11\,\mathrm{MeV}}{E}\sqrt{\frac{T}{X_0}}
\end{equation}
(Note that the coefficient 11 MeV here is appreciably smaller than
estimate
$\sqrt{\frac{4\pi}{\alpha}}\frac{m}{\sqrt2}=\frac{21.2\,\mathrm{MeV}}{\sqrt2}\simeq14.8$
MeV often used within the simplest leading logarithm approximation).

Inserting into Eq.~(\ref{sigma-am}) $T\sim R\theta_c$, and dividing
by Eq.~(\ref{aver-r-dot}), we arrive at a requirement
\begin{subequations}\label{eqs40}
\begin{equation}\label{mult-is-small}
    \frac{\sigma_{\mathrm{am}}(R\theta_c)}{\left\langle\dot{r}_{n_{\max}}\right\rangle}
    \simeq
    R\frac{11\,\mathrm{MeV}}{E}\sqrt{\frac{\theta_c}{X_0d}}
    \ll1,
\end{equation}
which may be viewed as a restriction on the crystal bending radius:
\begin{equation}\label{RllRmult}
    R\ll R_{\mathrm{mult}}(E)\qquad (\mathrm{coherent\, dynamics\, dominance}),
\end{equation}
\end{subequations}
if we introduce
\begin{eqnarray}\label{Rmult}
    R_{\mathrm{mult}}(E)=\frac{E}{11\,\mathrm{MeV}}\sqrt{\frac{X_0d}{\theta_c}}=\mathrm{m}\left(\frac{E}{18\,\mathrm{GeV}}\right)^{5/4}.\\
    (\text{positively charged particles})\qquad\qquad \nonumber
\end{eqnarray}
(We evaluate $\theta_c=\sqrt{2V_0/E}$ with $V_0=22.7$ eV for
Si(110)). From Eq.~(\ref{Rmult}), (\ref{Rc}) we note that at
relativistic particle energies definitely $R_{\mathrm{mult}}\gg
4R_c$. Hence, there exists a range of crystal curvatures $4R_c<
R<R_{\mathrm{mult}}$ covering the practically most relevant cases
$R>4R_c$, and at the same time not subject to strong multiple
scattering.

It is also instructive to compare the restriction
$R<R_{\mathrm{mult}}$ with that stemming from the crystal finite
thickness:
\begin{equation}\label{finite-length}
    R\ll L/\theta_c\sim100\,\mathrm{m}.
\end{equation}
Limitation (\ref{eqs40}) will be more restrictive than
(\ref{finite-length}) if
$\frac{L}{\mathrm{mm}}\gg\left(\frac{E}{1.75\,\mathrm{TeV}}\right)^{3/4}$.
Hence, at not very high energies and not very thin crystals, such as
they are nowadays, boundary effects on the volume reflection are
inessential. In principle, the boundary condition effect is to
reduce $\left\langle\theta_{\mathrm{defl}}\right\rangle$, but the
multiple scattering can manage that alone -- as we will discuss
below.

\subsection{Suppression of the mean angle of volume reflection by multiple scattering}

The effect of multiple scattering on
$\left\langle\theta_{\mathrm{defl}}\right\rangle$, which is needed
for input in Eqs.~(\ref{DL2}, \ref{DL-neg}), in the first order must
vanish, due to the symmetry of the incoherent scattering in the
scattering angle sign. Hence,
$\left\langle\theta_{\mathrm{defl}}\right\rangle$ is also a
convenient variable for representing the onset of multiple
scattering non-linearity.

In general, the strong multiple scattering manifests itself as
follows. Once condition (\ref{RllRmult}) is violated, the multiple
scattering can affect the particle trajectory in the last volume
reflection interval. At still larger $R$, it may affect the
trajectory even before the last interval is reached. But since the
non-zero mean deflection angle effect receives significant
contribution from the last interval, if the particle does not reach
it in due way, the volume reflection may be effectively terminated.
This must be definitely true for those particles which scatter
outwards in $r$, while those scattered inwards may still have a
chance to volume reflect later. Thus, at $R>R_{\mathrm{mult}}$ the
value of $\left\langle\theta_{\mathrm{defl}}\right\rangle$ is from
general reasons expected to decrease. Eventually, the particle net
deflection ought to vanish as the crystal straightens out. But the
rate of the decrease can not be assessed without a proper
calculation, or reference to experimental data.

On the other hand, it is well known that in the region $R<R_{
\mathrm{mult}}$ function
$\left\langle\theta_{\mathrm{defl}}\right\rangle(R)$ grows with the
increase of $R$. There exists a definite theoretical prediction for
the mean angle of volume reflection in a pure continuous potential
(applying harmonic approximation to the latter, which works well in
(110) orientation for silicon) \cite{ref:Bond-VR}. For positive
particles the formula reads\footnote{To see the correspondence with
Eq.~(73) of \cite{ref:Bond-VR}, note that
$\frac{d}{\theta_c^2}\approx2R_c$. But for a realistic potential,
when this equality is violated by $\simeq20\%$, as we argued in
\cite{ref:Bond-VR}, it is more accurate to use in the
$\frac1R$-correction term $\frac{d}{\theta_c^2}$ rather than $2R_c$.
}
\begin{equation}\label{thetavr}
    \left\langle\theta_{\mathrm{defl}}\right\rangle_{\mathrm{harm}}\approx\theta_{\lim}\left(1-\frac{d}{\theta_c^2R}\right),\qquad \theta_{\lim}=\frac\pi2\theta_c ,
\end{equation}
\begin{equation}\label{}
    (4R_c<R<R_{\mathrm{mult}},\quad \mathrm{positive\,
    particles}).
\end{equation}

Since function $\left\langle\theta_{\mathrm{defl}}\right\rangle(R)$
decreases to the both sides of $R_{\mathrm{mult}}$, it must achieve
a maximum somewhere around $R_{\mathrm{mult}}$ -- a prominent
feature to look for in computer simulation and experiments.

\subsection{Scaling law at large $R$}

\begin{figure}
\includegraphics{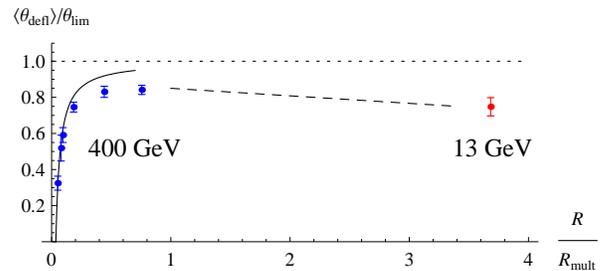}
\caption{\label{fig:theta-refl-exper} Dependence of the mean volume
reflection angle for positively charghed particles on the crystal
curvature. Blue points -- experimental data for protons at 400
GeV\cite{ref:Scand-R-dep}. Solid curve -- theoretical prediction
(\ref{thetavr}) taking into account only dynamics in the pure
continuous potential. Red points -- measurement for positively and
negatively charged particles at 13 GeV \cite{ref:Hasan}. Dashed line
-- interpolation.}
\end{figure}

At the present state of affairs, the data on volume reflection are
not very abundant, particularly at $R>R_{\text{mult}}$. For
positively charged particles and orientation (110), in the range
$R>R_{\text{mult}}$ there is only one experimental point -- at
$E=13$ GeV and $R=2.4$ m, which corresponds to
$R\simeq3.6R_{\text{mult}}$. On the other hand, at
$R<R_{\text{mult}}$ a relatively detailed experiment was performed
at $E=400$ GeV \cite{ref:Scand-R-dep}. To link these two
experiments, one may suggest that at $R>R_{\text{mult}}\gg R_c$,
scale $R_c$ drops out, and so
$\left\langle\theta_{\mathrm{defl}}\right\rangle/\theta_{\lim}$
becomes a function of the ratio $R/R_{\mathrm{mult}}(E)$ only. Then,
in the plot of
$\left\langle\theta_{\mathrm{defl}}\right\rangle/\theta_{\lim}$ vs.
$R/R_{\text{mult}}$ data at 13 GeV may be used to continue the
dependence at 400 GeV (see Fig.~\ref{fig:theta-refl-exper}). Its
behavior, slowly decreasing with $R$ beyond $R_{\text{mult}}$, is
close to that obtained in simulation \cite{ref:Tar-Scand} at a
single energy $400$ GeV \footnote{Though, in \cite{ref:Tar-Scand}
only $R$ values up to $\approx 1.5 R_{\text{mult}}$ were probed.}.
Thus, our expression (\ref{Rmult}) for $R_{\text{mult}}(E)$ for
positive particles is not unreasonable, and so is the scaling law
hypothesis.

For negatively charged particles, it is unobvious whether formula
(\ref{Rmult}) is applicable as well. Even if not, there must exist a
similar expression for $R_{\text{mult}}$ for negative particles,
monotonously increasing with $E$. However, there are experimental
indications \cite{ref:Hasan,ref:Scand-neg} that the volume
reflection pattern at the same energy for positively and negatively
charged particles strongly differs. For negative particles with the
decrease of energy, and hence increase of $R/R_{\text{mult}}$, the
mean deflection angle rapidly diminishes. We will not speculate
about this behavior until more detailed evidence arrives.

\section{Comparison with experiment for inelastic nuclear
interaction probability}\label{sec:Inel}

We are finally in a position to test predictions of
Sec.~\ref{sec:DeltaL} against the available experimental data. The
most straightforward check is supposed to be with the results of
experiments on inelastic nuclear scattering. At present, the only
such experiment is that with 400 GeV protons using a $L=2$ mm thick
silicon crystal at a single value of the crystal bending radius
$R=10$ m \cite{ref:Scand}.

\begin{figure}
\includegraphics[width=17pc]{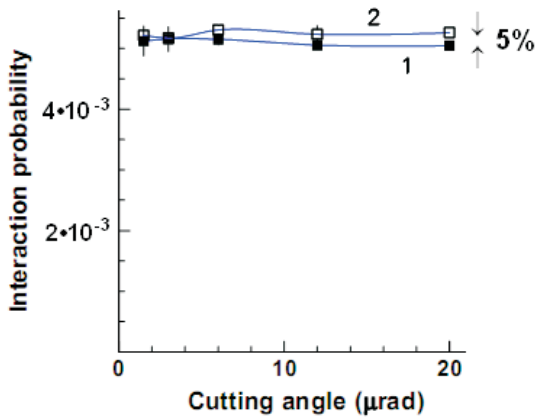}
\caption{\label{fig:5percent} (adapted from \cite{ref:Scand}).
Dependencies of the inelastic nuclear interaction probability of 400
GeV protons in the $R=10$ m crystal on the cutting angle of the
incident beam: (1) -- for `amorphous orientation', (2) -- for the
case of volume reflection.}
\end{figure}

The results of this experiment are displayed in
Fig.~\ref{fig:5percent} (adapted from \cite{ref:Scand}). In fact,
the inelastic nuclear interaction rate was measured for varying
cutting angles $\delta\theta_0$. When the cutting angle was
sufficiently large (which ought to correspond to perfect averaging
over $b$ or $E_\perp$), the measured relative difference was about
constant, on the level
\begin{equation}\label{5percent}
    \frac{\left\langle\Delta L\right\rangle}{L}\approx(5\pm2)\%.
\end{equation}

Our prediction, using the experimentally determined mean value
$\left\langle\theta_{\mathrm{defl}}\right\rangle_{\mathrm{exp}}=13.35\,\mu$rad
at the given curvature $1/R=0.1\mathrm{m}^{-1}$ amounts
to\footnote{If instead one employs formula (\ref{thetavr}), it
yields a somewhat larger
$\left\langle\theta_{\mathrm{defl}}\right\rangle$, and
correspondingly somewhat poorer agreement with the experiment.}
\begin{equation}\label{6.5}
    \frac{\left\langle\Delta L\right\rangle}{L}=\frac{R\left\langle\theta_{\mathrm{defl}}\right\rangle_{\mathrm{exp}}}{L}=6.67\%.
\end{equation}
The theoretical accuracy of prediction (\ref{6.5}) can be estimated
as $\mathcal{O}(R/R_{\mathrm{mult}})$ (see Eq.~(\ref{eqs40})).
Substituting here $R=10$ m, and
\[
R_{\text{mult}}(400\,\mathrm{GeV})\approx 50\text{ m},
\]
we infer $R/R_{\mathrm{mult}}\approx1/5$. That is commensurable with
the relative difference between our theory and the experiment
(Eqs.~(\ref{5percent}) and (\ref{6.5})). Needless to say, the
approximate agreement of one number does not conclusively validate
the theory. Measurements at different crystal thicknesses and
bending radii would be worthwhile.

A curious feature in Fig.~\ref{fig:5percent} is that at small
cutting angles ($\lesssim2\,\mu$rad) the difference between the rate
of inelastic nuclear interactions at volume reflection and in an
amorphous case seems to depart from a constant and actually vanish,
although experimental errors in this region are too high for an
unambiguous conclusion. In principle, some sensitivity to
$\delta\theta_0$ could emerge due to imperfect averaging and to the
impact of the second term in Eq.~(\ref{DelL}). To check this
possibility, let us estimate the range of variation of the argument
$\eta$ under the variation of $\theta_0$ in an interval
$\delta\theta_0\sim2\,\mu$rad. From Eq.~(\ref{var-theta0}) we get
\begin{equation}\label{}
    \delta\eta=\frac
    Rd\theta_0\delta\theta_0\sim\frac{L}{2d}\delta\theta_0\sim 10.
\end{equation}
Apparently, this number is still much greater than the unit interval
of variable $\eta$ definition, hence with the variation of
$\theta_0$ in that range, $\eta$ actually scans its definition
interval several times, i. e. cutting angles down to $\sim1\,\mu$rad
still can have no significant impact on $\Delta L$. So, we can only
attribute the small-cutting-amgle fluctuation of $\left\langle\Delta
L\right\rangle$ to enhanced experimental errors reflecting the
difficulty of achieving such small cutting angles.

\section{Angular spread due to elastic multiple
scattering}\label{sec:spread}

Registration of inelastic nuclear scattering discussed in the
previous Section requires dedicated instrumentation like beam loss
monitors. But even in a minimal beam deflection setup, the nuclear
interactions shall manifest themselves through an angular broadening
of the final beam. A complication here arises because the broadening
receives an additional contribution from the impact parameter
dependence of the deflection angle, even in a pure continuous
potential. In fact, the latter contribution is anisotropic, but the
beam spread transverse to the direction of deflection is rarely
measured, so in the published experimental data on the beam
dispersion in the direction of deflection they contribute together.
As usual, the main source of broadening is away from the volume
reflection region, but we suppose it to be subtractable. The
problem, again, is to explain the difference from the amorphous
orientation, and the difference between positive and negative
particles.

The proper subtraction is possible if the kinetics of the particle
passage through the crystal is decomposed into three distinct
stages: pure incoherent multiple scattering upstream the volume
reflection region (where the beam acquires Gaussian shape), pure
dynamical broadening in the volume reflection region, and pure
incoherent multiple scattering downstream of it. In terms of the
corresponding angular distribution functions that expresses as
\begin{equation}\label{ang-convol0}
    \frac{dw}{d\theta}=\int d\theta_2\frac{e^{-\frac{(\theta-\theta_2)^2}{2\sigma_2^2}}}{\sqrt{2\pi}\sigma_2}
    \int
    d\theta_1\frac{dw_{\mathrm{coh}}\left(\theta_2-\theta_1\right)}{d\left(\theta_2-\theta_1\right)}\frac{e^{-\frac{\theta_1^2}{2\sigma_1^2}}}{\sqrt{2\pi}\sigma_1},
\end{equation}
where we assumed the upstream and downstream incoherent scattering
to be purely Gaussian, and $dw_{\mathrm{coh}}$ is the angular
distribution function in a pure continuous potential (slightly
averaged over the particle incidence angles). The small portion of
multiple Coulomb scattering within the volume reflection region may
be included either in the upstream or downstream scattering piece,
as long as it is small, and thus additive. Moreover, if in
(\ref{ang-convol0}) we change the integration variables, one
integration can be taken, with the result
\begin{equation}\label{ang-convol}
    \frac{dw}{d\theta}=\int d\alpha
    \frac{dw_{\mathrm{coh}}(\alpha)}{d\alpha}\frac1{\sqrt{2\pi\left(\sigma_1^2+\sigma_2^2\right)}}e^{-\frac{(\theta-\alpha)^2}{2\left(\sigma_1^2+\sigma_2^2\right)}}
\end{equation}
depending only on the sum $\sigma_1^2+\sigma_2^2$. Obviously, the
latter sum must be equated to $\sigma^2_{\mathrm{am}}(L+\Delta L)$,
and thereby the precise positions of the boundaries separating the
different kinetic regions prove to be inessential.

Examining Eq.~(\ref{ang-convol}), one should realize that for real
crystals, usually, the width of $dw_{\mathrm{coh}}/d\alpha$ is
smaller than $\sigma^2_{\mathrm{am}}$, whereby the resulting angular
distribution becomes close to a Gaussian, anyway. So, it is
described essentially in terms of two moments:
\begin{equation}\label{}
    \left\langle\theta\right\rangle=\int
    d\theta\theta\frac{dw}{d\theta}\qquad \left(\int
    d\theta\frac{dw}{d\theta}=1\right),
\end{equation}
and
\begin{equation}\label{}
    \sigma^2=\int
    d\theta\left(\theta-\left\langle\theta\right\rangle\right)^2\frac{dw}{d\theta}.
\end{equation}
The mean value $\left\langle\theta\right\rangle$ only receives
contribution from $dw_{\mathrm{coh}}/d\theta$:
\begin{equation}\label{}
    \left\langle\theta\right\rangle=\int
    d\theta\theta\frac{dw_{\mathrm{coh}}}{d\theta}
\end{equation}
(as was implied in Sec.~\ref{sec:Rmult}), while when we evaluate
$\sigma^2$ from Eq.~(\ref{ang-convol}), the coherent and incoherent
contributions to it appear to be just additive:
\begin{eqnarray}\label{sumsigmas}
    {\sigma}^2&=&\sigma^2_{\mathrm{am}}(L+\Delta L)+\sigma^2_{\mathrm{coh}}\nonumber\\
    &\equiv&\sigma^2_{\mathrm{am}}(L)+\sigma^2_{\mathrm{am}}(R\left\langle\theta_{\mathrm{defl}}\right\rangle)+\sigma^2_{\mathrm{coh}},
\end{eqnarray}
with
\begin{equation}\label{}
    \sigma^2_{\mathrm{coh}}=\int
    d\theta\left(\theta-\left\langle\theta\right\rangle\right)^2\frac{dw_{\mathrm{coh}}}{d\theta}.
\end{equation}

Angular distribution $dw_{\mathrm{coh}}/d\theta$ was evaluated in
\cite{ref:Bond-VR} in the model of harmonic continuous potential
between (110) silicon crystallographic planes. It has some
differences, for positive and negative particles, as does the
nuclear interaction rate calculated in Sec.~\ref{sec:DeltaL}. Let us
begin with the case of positive particles, to which most of the data
refer.

\subsection{Positively charged particles}

For positive particles, at $R>4R_c$ the coherent part of the angular
distribution looks as (see Eq.~(72) of
\cite{ref:Bond-VR})\footnote{In paper \cite{ref:Bond-VR} the fianl
beam angular distribution was described in terms of
$d\lambda/d\theta$, the differential cross-section. But obviously,
dividing that quantity by $d$, we obtain the normalized probability
distribution
$\frac{dw_{\mathrm{coh}}}{d\theta}=\frac{1}{d}\frac{d\lambda}{d\theta}$,
$\int d\theta \frac{dw_{\mathrm{coh}}}{d\theta}=1$ dealt with in the
present paper.}
\begin{equation}\label{}
    \frac{dw_{\mathrm{coh}}}{d\theta}=\frac{R\theta_c}{\pi d}\Theta\left(\frac{\pi
    d}{2R\theta_c }-|\theta-\left\langle\theta\right\rangle|\right),
\end{equation}
$\Theta$ being the Heavyside unit step function, and
$\left\langle\theta\right\rangle$ being given by
Eq.~(\ref{thetavr}). The corresponding $\sigma_{\mathrm{coh}}$
ensues as
\begin{equation}\label{sigma2vr}
    \sigma_{\mathrm{coh}}\approx\frac{\pi}{2\sqrt3
    \theta_c}\frac{d}{R},
\end{equation}
notably being inversely proportional to the crystal bending radius.
For a realistic continuous potential the numerical coefficient in
(\ref{sigma2vr}) may slightly differ, but that is not crucial for
the following estimates.

Measurements of total ${\sigma}^2$ for 400 GeV protons interacting
with a (110) silicon crystal were carried out in experiment
\cite{ref:Scand-R-dep}. There, in order to get access to the
intrinsic volume reflection angular divergence
$\sigma_{\mathrm{coh}}$, the difference
\begin{equation}\label{}
    \sigma^2-\sigma^2_{\mathrm{am}}(L)=\bar\sigma^2_{\mathrm{v.r.}}
\end{equation}
was evaluated. From Eq.~(\ref{sumsigmas}) we see, however, that it
differs from pure $\sigma_{\mathrm{coh}}$:
\begin{equation}\label{bb}
    \bar{\sigma}_{\mathrm{v.r.}}=\sqrt{\sigma^2-\sigma^2_{\mathrm{am}}}=\sqrt{\sigma^2_{\mathrm{coh}}+\sigma^2_{\mathrm{am}}(R\left\langle\theta_{\mathrm{defl}}\right\rangle)},
\end{equation}
\[
    \bar{\sigma}_{\mathrm{v.r.}}\neq {\sigma}_{\mathrm{coh}}.
\]

Assuming condition (\ref{mult-is-small}) to hold, we may insert
explicit theoretical expressions (\ref{sigma-am}), (\ref{sigma2vr}),
(\ref{thetavr}) into Eq.~(\ref{bb}), which leads to a non-scaling
$R$-dependence of the measured quantity
$\bar{\sigma}_{\mathrm{v.r.}}$:
\begin{eqnarray}\label{barsigma}
\bar{\sigma}_{\mathrm{v.r.}}=\sqrt{\frac{\pi^2}{12\theta_c^2}\frac{d^2}{R^2}+\frac{\pi\theta_c}{2}\left(\frac{11\,\mathrm{MeV}}{E}\right)^2\frac{R-d/\theta_c^2}{X_0}}\\
(R<R_{\mathrm{mult}}).\qquad\qquad\qquad\quad\nonumber
\end{eqnarray}
The most characteristic feature of function (\ref{barsigma}) is the
existence of a minimum. The minimum location is found by equating to
zero the derivative of the radicand with respect to $R$:
\begin{equation}\label{Rmin}
    R_{*}(E)=\frac1{\theta_c}\sqrt[3]{\frac{\pi}3
    X_0d^2}\left(\frac{E}{11\,\mathrm{MeV}}\right)^{2/3}\simeq\left(\frac{E}{38\,\mathrm{GeV}}\right)^{7/6}\mathrm{m}.
\end{equation}
The physical meaning of $R_{*}$ is not vastly different from that of
$R_{\mathrm{mult}}$ -- it marks the scale of $R$ where the multiple
scattering compares with coherent deflection angles, with the
proviso that $R_{\mathrm{mult}}$ is derived from generic reasoning
in terms of the particle trajectory, while $R_{*}$ -- in terms of
specific contributions to the beam broadening in the crystal.
However, the actual expressions for $R_{\mathrm{mult}}$ and $R_{*}$
differ; moreover, their ratio
\begin{equation}\label{ratio}
    \frac{R_{\mathrm{mult}}}{R_{*}}=\left(\frac3\pi
    \frac{E}{11\,\mathrm{MeV}}\right)^{1/3}\!\left(\!\frac{X_0}{d}\!\right)^{1/6}\!\sqrt{\frac{\theta_c}2}\simeq  \left(\frac{E}{50\, \mathrm{MeV}}\right)^{1/12}
\end{equation}
even depends on the particle energy, albeit pretty weakly. At
ultra-relativistic energies ratio (\ref{ratio}) is $>1.5$,
justifying the use of Eq.~(\ref{barsigma}), but up to LHC energies
it does not exceed $3$. The minimal value of
$\bar\sigma_{\mathrm{v.r.}}$
\begin{equation}\label{sigma-min}
    \bar{\sigma}_{\min}=\bar{\sigma}_{\mathrm{v.r.}}(R=R_{*})\simeq
    \left(\frac{1\,\mathrm{keV}}{E}\right)^{2/3}\
\end{equation}
sets the scale of $\bar{\sigma}_{\mathrm{v.r.}}$ at $R\gtrsim R_*$,
because beyond $R_*$ function (\ref{barsigma}) varies slowly.

\begin{figure}
\includegraphics{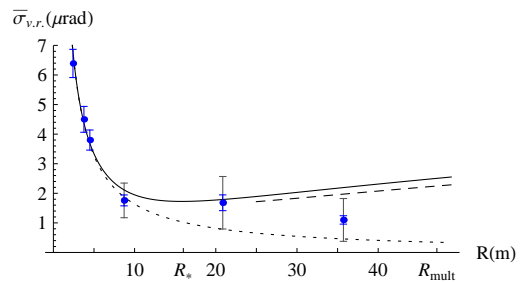}
\caption{\label{fig:exper-vs-theor} Subtracted final beam angular
width vs. the crystal bending radius, for $E=400$ GeV protons in a
$L=2$ mm silicon crystal. Solid curve -- theoretical prediction
(Eq.~(\ref{barsigma})). Dashed curve -- with experimental
$\left\langle\theta_{\mathrm{defl}}\right\rangle$. Dotted curve --
pure $\sigma_{\mathrm{coh}}$. Blue points -- experimental data
\cite{ref:Scand-R-dep}. Gray error bars -- evaluated by
Eq.~(\ref{propag}).}
\end{figure}

The available data at $E=400$ GeV (presented in
Fig.~\ref{fig:exper-vs-theor}) do not reach beyond
$R_{\mathrm{mult}}$, but do reach well beyond $R_*$, which according
to Eq.~(\ref{Rmin}) equals
\begin{equation}\label{}
    R_*(400\text{ GeV})\approx16\,\mathrm{m}.
\end{equation}
Around $R_*$ the data show a flattening of the $R$-dependence, but
the point $R=35.71$ m closest to $R_{\mathrm{mult}}$ seems to resume
the decrease of $\bar\sigma_{\mathrm{v.r.}}$, which is rather
unexpected. It must be noted that at this value of $R$ the final
beam shape qualitatively changes -- there develops a thrust in the
elsewhere Gaussian profile, extending to the side of the crystal
bending, though not as far as to the channeling angle $L/2R$ (see in
\cite{ref:Scand-R-dep} Fig.~3, left panel). That component is not
associated with volume reflection, and is conventionally attributed
to volume capture, but still remains a poorly understood phenomenon
(see, e.g., \cite{ref:Tar-Scand,ref:VCapture}). Besides particle
capture to the channeling regime, one can in principle imagine other
mechanisms of particle drag to the side of the crystal bending, e.
g., being due to detention of a fraction of protons on the
\emph{top} of a curved potential barrier, containing an area of
strong multiple scattering. In either case, it is not very
surprising that an additional component arises as $R$ approaches
$R_{\mathrm{mult}}$.

Concerning the observed reduction of $\bar\sigma_{\mathrm{v.r.}}$
compared to the theoretical prediction (\ref{barsigma}) at $R=35.71$
m, one might say that in view of the final beam profile
non-Gaussianity, as well as the ambiguity of the separation of
volume reflected and volume captured fractions, our description
neglecting the volume capture is invalidated as a whole, at least
for description of $\bar\sigma_{\mathrm{v.r.}}$. On the other hand,
the volume captured fraction comprises only 6\% of the particles,
and the volume reflected fraction is still Gaussian, whereas the
discrepancy with the theory by a factor of $\sim2$ appears to be
exceedingly large.

In the search of an explanation to the encountered discrepancy, we
must pay attention to the fact that at evaluation of a difference
between two close quantities the experimental errors enhance.
Evaluating propagation of errors for $\bar\sigma_{\mathrm{v.r.}}$
defined by Eq.~(\ref{bb}), we strictly derive
\begin{equation}\label{}
    \sqrt{(\sigma\pm\delta\sigma)^2-(\sigma_{\mathrm{am}}\pm\delta\sigma_{\mathrm{am}})^2}
    =\bar\sigma_{\mathrm{v.r.}}\pm\delta\bar\sigma_{\mathrm{v.r.}},
\end{equation}
with
\begin{subequations}\label{propag}
\begin{eqnarray}
    \delta\bar\sigma_{\mathrm{v.r.}}&=&\frac{\sqrt{(\sigma\delta\sigma)^2+(\sigma_{\mathrm{am}}\delta\sigma_{\mathrm{am}})^2}}{\bar\sigma_{\mathrm{v.r.}}}\\
    &\approx&
    \frac
{\sigma_{\mathrm{am}}} {\bar\sigma_{\mathrm{v.r.}}}
    \sqrt{(\delta\sigma)^2+(\delta\sigma_{\mathrm{am}})^2}\qquad (\mathrm{if}\,\sigma_{\mathrm{am}} >
    \bar\sigma_{\mathrm{v.r.}}).\nonumber\\
    \label{73b}
\end{eqnarray}
\end{subequations}
Now, for the point at $R=35.71$ m  the prefactor
$\sigma_{\mathrm{am}}/\bar\sigma_{\mathrm{v.r.}}$ in Eq.~(\ref{73b})
amounts $\sim5$, but the experimental error for
$\bar\sigma_{\mathrm{v.r.}}$ is quoted to be as small as
$\delta\bar\sigma_{\mathrm{v.r.}}=\sqrt{0.11^2+0.09^2}\mu\mathrm{rad}=0.14\mu\mathrm{rad}$,
being comparable to
$\delta\sigma_{\mathrm{am}}=0.11\mu\mathrm{rad}$. Thus, probably,
the error propagation factor of 5 yet needs to be added for this
point, and for consistency -- for the preceding two points as
well\footnote{For the rest of the points that correction is of
little consequence, since there
$\sigma_{\mathrm{am}}/\bar\sigma_{\mathrm{v.r.}}\sim1$.}. The error
bars modified in this way are displayed in
Fig.~\ref{fig:exper-vs-theor} by gray color. The theoretical curve,
in turn, may shift somewhat lower (see the dashed curve in
Fig.~\ref{fig:exper-vs-theor}) if instead of $\theta_{\lim}$ one
uses the empirically determined
$\left\langle\theta_{\mathrm{defl}}\right\rangle$, as in
Fig.~\ref{fig:theta-refl-exper}. Therewith, the discrepancy between
experiment and the theory appears to be just within one standard
deviation. Formally, it may even comply with $\sigma_{\mathrm{coh}}$
as well (the dotted line in Fig.~(\ref{fig:exper-vs-theor})), but in
the light of the foregoing analysis, that seems unlikely.

\subsection{Negatively charged particles}

Experimental data for negative particles at present are too scarce
to allow phenomenological investigations of
$\bar\sigma^2_{\mathrm{v.r.}}$, so we will describe for them the
situation in general terms.

First of all, for negatively charged particles the expression for
$\sigma_{\mathrm{coh}}(R)$ somewhat differs, though its
$R$-dependence remains close to $1/R$, up to logarithmic factors.
Secondly, in this case we have
$\Delta\sigma^2_{\mathrm{am}}\propto\left\langle\Delta
L\right\rangle<0$. Therefore, the expression for
$\bar\sigma^2_{\mathrm{v.r.}}$ for negative particles is similar to
the radicand of Eq.~(\ref{barsigma}), only with a negative
coefficient at the second term. That implies that for negative
particles $\bar\sigma^2_{\mathrm{v.r.}}$ changes sign and becomes
\emph{negative} for sufficiently large $R$. That is the salient
feature of the final beam angular distribution for negative
particles, which would be interesting to verify experimentally.

The second remark is that since $\sigma_{\mathrm{coh}}$ for
positively and for negatively charged particles differ, in general
it is not as straightforward to compare the broadenings for positive
and negative particles, as that was the case with the rate of
inelastic nuclear interactions. However, in the region $R>R_c$ where
$\sigma_{\mathrm{coh}}$ gets relatively small, that must already be
possible. The simplest way of pinning down $\sigma_{\mathrm{coh}}$,
though, is to measure both angular beam broadening components
perpendicular and parallel to the family of the active atomic
planes.

\section{Summary and conclusions}\label{sec:Summary}

In the present article we were concerned with the problem of
incoherent nuclear scattering at volume reflection. This problem
belongs to the category of combined potential and stochastic motion,
which is complicated even in the radial 1d case, hampering advances
without a numerical simulation. However, it turned out that small
(perturbatively tractable) multiple scattering limit covers the
practically most important region $R_c\ll R< R_{\mathrm{mult}}(E)$,
with $R_{\mathrm{mult}}(E)$ defined by Eq.~(\ref{Rmult}). There, by
analytic means we attained a simple relation that the difference
between the probability of nuclear interactions of a proton in a
bent crystal and in an amorphous target is proportional to the
product of the beam mean deflection angle and the crystal bending
radius (Eqs.~(\ref{via}), (\ref{DL2}), (\ref{39})). That relation
exploits only the local character of nuclear interactions and the
periodicity of the atomic planes, and does not resort to any
parameterization for the continuous potential, such as the parabolic
approximation used in our earlier treatment of the volume reflection
\cite{ref:Bond-VR}.

Our perturbative prediction for positive particles was confronted
with the experimental data \cite{ref:Scand}, \cite{ref:Scand-R-dep},
both for the rate of inelastic nuclear interactions
(Fig.~\ref{fig:5percent}) and for the volume-reflected beam
broadening (Fig.~\ref{fig:exper-vs-theor}). The theoretical
predictions are in an encouraging agreement with the data, though
some unclear phenomena still remain on edges of measurement
intervals.

The utility of the predicted correction is twofold. Firstly, it may
be just a noticeable correction which needs account. With the trend
to decrease $L$ in order to reduce the background from useless
multiple Coulomb scattering, the relative significance of the
crystal curvature dependent nuclear scattering must increase.
Secondly, upon accumulation of our experience on behavior of volume
reflection at different experimental conditions it may grow into a
method of monitoring the state of the volume reflection dynamics
inside the crystal. The guiding principle here must be, basically,
that the greater the nuclear scattering difference, the more robust
the volume reflection dynamics.

In conclusion, let us point out that besides the domain
$R<R_{\mathrm{mult}}$ which matters most for the accelerator
applications, it is also of scientific interest to understand the
particle passage process under conditions $R\gtrsim
R_{\mathrm{mult}}$, where volume reflection merges with the volume
capture. In Secs.~\ref{sec:Rmult}, \ref{sec:spread} we tried to give
some qualitative insight into these issues. The results obtained in
the present article may also serve as a guide for more fundamental
computer simulations.

\end{document}